\documentclass[MSNbibl,nameyear,dvips]{arxstspdf}
\usepackage{flushend}
\usepackage{stfloats}

\volume{26}
\issue{2}
\pubyear{2011}
\firstpage{235}
\lastpage{237}
\doi{10.1214/11-STS331C}
\referstodoi{10.1214/10-STS331}

\makeatletter
\def\bsuffix #1{#1}
\makeatother

\begin{document}
\begin{frontmatter}

\title{Sampling from a Bayesian Menu}
\runtitle{Discussion}
\pdftitle{Discussion of Bayesian Models and Methods in
Public Policy and Government Settings by S. E. Fienberg}

\begin{aug}
\author{\fnms{Alan M.} \snm{Zaslavsky}\corref{}\ead[label=e1]{zaslavsk@hcp.med.harvard.edu}}
\runauthor{A. M. Zaslavsky}

\affiliation{Harvard University}

\address{Alan M. Zaslavsky is Professor of Health Care Policy
(Statistics), Department of Health Care Policy, Harvard Medical School 180 Longwood Ave.,
Boston, Massachusetts 02115, USA \printead{e1}.}

\end{aug}



\end{frontmatter}

I am pleased that Steve Fienberg's article opens a~discussion aimed at
broadening the scope of statistical methods applied to policy problems.
His \textit{mezes} platter of case studies whets the appetite for a deeper
study of these application areas.  My further thoughts largely center
on just what it means to say that the examples he gives (some quite
delicious, especially the aged wine of electoral projections) are
``Bayesian.''  Fienberg argues on a combination of intellectual and
historical grounds for a unitary view of Bayesian statistics, thus
bringing a broad range of statistical practice and applications under
the Bayesian awning.  Despite the advantages of such a~comprehensive
view, it is also useful to distinguish the components, both to clarify
their relationships and so consumers of methodology who are not
prepared to eat the entire \textit{prix fixe} dinner can still order off
the menu what suits their tastes and nutritional needs.  While
Fienberg's presentation emphasizes the inferential entr\'{e}e, the
assessment of posterior probabilities, it may help to detail the
offerings on the Bayesian menu:

 Main courses:
\begin{itemize}
\item A subjectivist understanding of probability, allow\-ing for
meaningful probability statements about singular events. \item
Comprehensive model specification, including
\begin{itemize}
\item Likelihoods.
\item Prior distributions.
\end{itemize}
\item Use of Bayes's theorem to ``turn the Bayesian crank,'' making
inferences about parameters (and possibly predictive statements about
unobserved or future populations).
\end{itemize}

Optional dishes:
\begin{itemize}
\item Subjective priors incorporating substantive prior beliefs. \item
Model selection by Bayesian methods; model mixing. \item Hierarchical
modeling.
\end{itemize}

Some of these dishes are commonly ordered \textit{a la carte}.  Obviously,
modeling is a central component of statistical practice for
statisticians of a variety of schools, although a non-Bayesian
generally has more leeway to introduce nonmodel-based procedures (such
as resampling methods) into the mix.  In particular, despite the
theoretical and historical connections Fienberg notes of hierarchical
modeling to Bayesian concepts of exchangeability, one need not be a
Bayesian to use hierarchical models, applying maximum likelihood
estimation at the top level, so-called Maximum Likelihood Empirical
Bayes (MLEB), or with some other non-Bayesian procedure.  Estimation
for level 2 parameters (``random effects'' for the frequentist) may
proceed using Bayes's law, or by appealing to completely non-Bayesian
arguments like BLUP (best linear unbiased prediction), thus eating the
Bayesian omelet while getting only the faintest whiff of the Bayesian
eggs.

Distaste for Bayesian statistical approaches in policy settings arises
at various points in this menu.  For the census, which each of the 435
members of the House of Representatives views through the lens of its
impact on his or her own district, any use of modeling aroused
immediate suspicion due to fears of manipulation of possibly arbitrary
model specifications.  Similar concerns contribute to the general
dominance of ``design-consistent'' classical survey sampling methods in
government statistics, even when ``model-assisted.''  It is noteworthy
that the statistical objections to using hierarchical models in
estimation of census undercount centered on the use of any regression
model that pooled information across states, not particularly on the
use of a hierarchical model (fully Bayesian or MLEB).  Finally, the
Supreme Court ruled in 1999 against any use of sampling for census
apportionment counts, even with estimation based on the purest of
``design-based'' principles of unbiased survey estimation,
citing\vadjust{\eject}
concerns of susceptibility to manipulation, or at least to controversy.
(Oddly enough, the deciding opinion by Justice O'Connor hinged largely
on interpretation of a grammatical construction in two apparently
conflicting sections of the Census Act, as well as the interpretation
of the constitutional phrase ``actual enumeration.'')  It is noteworthy
that nonstatistical details of census data-collection methodology that
might have equal or greater effects on outcomes, such as the nature of
the public awareness efforts or the number of in-person follow-up
attempts to mail nonrespondents, are rarely subject to the same degree
of scrutiny.

Nonetheless, I agree that the main philosophical objection to entering
the Bayesian restaurant at all concerns the choice of prior.  In this
regard, the distinction between ``objective'' and ``subjective'' Baye\-sian
approaches becomes significant.  As I understand the objective
approach, it does not require the analyst to be committed to a prior as
a representation of substantive prior beliefs, but only as a~generic
device that leads to Bayesian inferences with good frequency
properties, that is, one which generates calibrated probability
statements, in the spirit of Rubin (\citeyear{Rub84}).  Even improper priors, which
since they are not probability distributions cannot be regarded as
coherent statements of prior beliefs, are acceptable if they lead to
posterior distribution with good frequency properties over the desired
range in the hyperparameter space.  The analyst gains access to a
well-specified inferential approach with a well-developed set of
techniques for estimation of posterior distributions, and thus ``eats a
Bayesian omelet made with powdered Bayesian eggs''---perhaps not as
tasty a dish as an inference based on a more substantive prior, but
nourishing nonetheless.  I would place both the census and GOM
disability examples of Fienberg's article in this category.  In neither
of these cases do I see choice of a prior as a significant obstacle.
To give a fairly typical example, O'Malley and Zaslavsky (\citeyear{OMaZas08})
estimated a correlation matrix in a~multilevel model using several
default priors, comparing results from those that are flexible enough
to have desirable properties of near-invariance to scale.  As in Fay
and Herriott (\citeyear{FayHer79}), the likelihood at the lowest level of the model is
approximated by a non-Bayesian calculation without a~complete model for
the complex survey data structure.

The subjective Bayesian begins with an informative prior representing
substantive beliefs. Such beliefs might be based on expert consensus
(elicited directly from experts or drawn from a review of the
literature) or inferred from relevant prior data.  In the latter case,
the evidence might take the form of a likelihood for the previous data,
with parameters linked to those presently of interest through a
hierarchical model, possibly with default ``objective'' priors for
hyperparameters (or even estimated by MLEB, although the fully\ Bayesian
model more readily accommodates uncertainty about these parameters).
For example, we might regard a trial of a new drug as a priori part of
an exchangeable sequence (conditional on some covariates) of trials of
the same or comparable drugs.  (I~particularly enjoyed Fienberg's
exposition of the successes and tri\-bulations of such Bayesian
approaches at the Food and Drug Administration.)  The substance of the
(scientific and policy) debate over the prior then concerns the choice
of the ensemble of relevant previous trials and the specification of
the way in which the results are believed to relate to each other,
essentially recasting this part of the model as a Bayesian
meta-analysis.  Metahypotheses about how such evidence should be
combined might be evaluated in the long run by the same criteria of
goodness of fit and predictive validity as are used in any other model
selection problem.  Notably, many Bayesians favor such frequency
criteria in model selection (Rubin, \citeyear{Rub84}), departing from a purely
Bayesian paradigm; the latter might suggest relying on model averaging
among a number of a priori reasonable models, but this compounds the
problem of choosing and justifying a prior distribution.

Fienberg's climate change case study illustrates how a Bayesian
perspective offers a principled framework for combination of sources of
uncertainty.  A~simpler example of the same principle concerns
microsi\-mulation modeling of food stamp benefits (Zaslavsky and
Thurston, \citeyear{ZasThu95}; Thurston and Zaslavsky, \citeyear{ThuZas96}).  In these models,
records on individuals are processed by algorithms representing the
application of current program rules and proposed modifications to
calculate the impact of possible changes.  Uncertainties take a variety
of forms:  sampling variation in the underlying database, stochastic
simulation error, and uncertainty among alternative assumptions about
future macroeconomic conditions and about parameters of submodels used
to correct measurement error or to impute variables not observed in the
underlying surveys.  Nonsubjectivist views of probability offer no
coherent framework for combining these various forms of uncertainty.
From a Bayesian perspective, however, each is a~contributor to
posterior variation; variance components can be partitioned and
attributed to the various kinds of uncertainty by applying ANOVA to
results of a designed experiment in which the factors are
systematically manipulated.  The resulting estimates show which
uncertainties are most important for each estimand of\vadjust{\goodbreak} interest, and
therefore suggest how effort might be best directed to reduce
uncertainty by conducting additional simulations, obtaining more data,
or seeking more consensus on particular economic or modeling
assumptions.

In conclusion, there is too much at stake in current policy-making to
require it to rely on a single statistical philosophy.  While not
everything Fienberg describes is the exclusive property of Bayesians,
it may well be the case that only those methodologists whose training
gives them a taste for Bayesian perspectives (rather than an allergy to
them) will be prepared to apply these tools.  I applaud Fienberg for
demonstrating, under the general rubric of Bayesian statistics, how
modeling in general,\vadjust{\eject} hierarchical modeling in particular, and Bayesian
philosophical approaches can enrich the toolkit for policy analysis.\



\begin{thebibliography}{5}

\bibitem[\protect\citeauthoryear{Fay and Herriot}{1979}]{FayHer79}
\begin{barticle}[mr]
\bauthor{\bsnm{Fay},~\bfnm{Robert~E.}\binits{R.~E.} \bsuffix{III}} \AND
  \bauthor{\bsnm{Herriot},~\bfnm{Roger~A.}\binits{R.~A.}}
(\byear{1979}).
\btitle{Estimates of income for small places: An application of {J}ames--{S}tein
  procedures to census data}.
\bjournal{J. Amer. Statist. Assoc.}
\bvolume{74}
\bpages{269--277}.
\bid{issn={0003-1291}, mr={0548019}}
\end{barticle}
\endbibitem



\bibitem[\protect\citeauthoryear{O'Malley and Zaslavsky}{2008}]{OMaZas08}
\begin{barticle}[mr]
\bauthor{\bsnm{O'Malley},~\bfnm{A.~James}\binits{A.~J.}} \AND
  \bauthor{\bsnm{Zaslavsky},~\bfnm{Alan~M.}\binits{A.~M.}}
(\byear{2008}).
\btitle{Domain-level covariance analysis for multilevel survey data with
  structured nonresponse}.
\bjournal{J. Amer. Statist. Assoc.}
\bvolume{103}
\bpages{1405--1418}.
\bid{doi={10.1198/016214508000000724}, issn={0162-1459}, mr={2655721}}
\end{barticle}
\endbibitem

\bibitem[\protect\citeauthoryear{Rubin}{1984}]{Rub84}
\begin{barticle}[mr]
\bauthor{\bsnm{Rubin},~\bfnm{Donald~B.}\binits{D.~B.}}
(\byear{1984}).
\btitle{Bayesianly justifiable and relevant frequency calculations for the
  applied statistician}.
\bjournal{Ann. Statist.}
\bvolume{12}
\bpages{1151--1172}.
\bid{doi={10.1214/aos/1176346785}, issn={0090-5364}, mr={0760681}}
\end{barticle}
\endbibitem

\bibitem[\protect\citeauthoryear{Thurston and Zaslavsky}{1996}]{ThuZas96}
\begin{bincollection}[auto:STB|2011-03-03|12:04:44]
\bauthor{\bsnm{Thurston},~\bfnm{S.~W.}\binits{S.~W.}} \AND
  \bauthor{\bsnm{Zaslavsky},~\bfnm{A.~M.}\binits{A.~M.}}
(\byear{1996}).
\btitle{Variance estimation in microsimulation models of the food stamp
  program}.
In \bbooktitle{ASA Proceedings of the Social Statistics Section}
\bpages{4--9}.
\bpublisher{Amer. Statist. Assoc.}, \baddress{Alexandria, VA}.
\end{bincollection}
\endbibitem

\bibitem[\protect\citeauthoryear{Zaslavsky and Thurston}{1995}]{ZasThu95}
\begin{bmisc}[auto:STB|2011-03-03|12:04:44]
\bauthor{\bsnm{Zaslavsky},~\bfnm{A.~M.}\binits{A.~M.}} \AND
  \bauthor{\bsnm{Thurston},~\bfnm{S.~W.}\binits{S.~W.}}
(\byear{1995}).
\bhowpublished{Error analysis of food stamp microsimulation models: Further results.
  In \textit{ASA Proceedings of the Social Statistics Section} 151--156.
  Amer. Statist. Assoc., Alexandria, VA}.
\end{bmisc}
\endbibitem
\vspace*{-2pt}
\end{thebibliography}
\end{document}